\documentclass[twocolumn,aps,prd,superscriptaddress,nofootinbib,floatfix]{revtex4-1}

\usepackage{graphicx,multirow}
\usepackage{amsmath,amsfonts}
\usepackage{hyperref,breakurl}
\usepackage{bm, xcolor}
\usepackage{slashed}

\newcommand{\nn}{\nonumber}
\newcommand{\be}{\begin{eqnarray}}
\newcommand{\ee}{\end{eqnarray}}
\newcommand{\beq}{\begin{equation}}
\newcommand{\eeq}{\end{equation}}
\newcommand{\beqa}{\begin{eqnarray}}
\newcommand{\eeqa}{\end{eqnarray}}
\newcommand{\TeV}{\text{ TeV}}
\newcommand{\GeV}{\text{ GeV}}

\newcommand{\lqcd}{\ensuremath{\Lambda_\text{QCD}}}

\newcommand{\eq}[1]{Eq.~\eqref{eq:#1}}

\renewcommand{\sec}[1]{Sec.~\ref{sec:#1}}
\newcommand{\ssec}[1]{Sec.~\ref{ssec:#1}}

\newcommand{\fig}[1]{Fig.~\ref{fig:#1}}

\newcommand{\tab}[1]{Table~\ref{tab:#1}}

\newcommand{\ds}{\displaystyle}
\def\d{{\rm d}}
\def\OMIT#1{{}}

\newcommand{\Bbar}{\,\overline{\!B}{}}
\newcommand{\Dbar}{\,\overline{\!D}{}}
\newcommand{\Kbar}{\,\overline{\!K}{}}
\def\B0bar{\Bbar{}^0}
\def\D0bar{\Dbar{}^0}
\def\K0bar{\Kbar{}^0}

\def\GeV{\text{ GeV}}
\def\TeV{\text{ TeV}}

\newcommand{\bra}[1]{\langle #1 |}
\newcommand{\ket}[1]{| #1 \rangle}

\newcommand{\repone}{\mathbf{1}}
\newcommand{\reptwo}{\mathbf{2}}
\newcommand{\repthr}{\mathbf{3}}
\newcommand{\arepthr}{\bar{\mathbf{3}}}

\definecolor{red}{rgb}{1,0,0}

\begin{document}

\title{\boldmath Flavor models for $\bar{B} \to D^{(*)}\tau\bar{\nu}$}

\author{Marat Freytsis}
\affiliation{Department of Physics, Harvard University, Cambridge MA, 02138}

\author{Zoltan Ligeti}
\affiliation{Ernest Orlando Lawrence Berkeley National Laboratory,
University of California, Berkeley, CA 94720}

\author{Joshua T.\ Ruderman}
\affiliation{Center for Cosmology and Particle Physics, Department of Physics, New York University, New York, NY 10003}

\begin{abstract}

The ratio of the measured $\bar B\to D^{(*)}\ell\bar{\nu}$ decay rates for $\ell
= \tau$ vs.\ $e,\mu$ deviate from the Standard Model (SM) by about $4\sigma$. 
We show that the data are in tension with the SM, independent of form factor
calculations, and we update the SM prediction for $\mathcal{B}(B\to
X_c\tau\bar{\nu}) / \mathcal{B}(B\to X_c \ell\bar{\nu})$.  We classify the
operators that can accommodate the measured central values, as well as their UV
completions. We identify models with leptoquark mediators that are minimally
flavor violating in the quark sector, and are minimally flavor violating or
$\tau$-aligned in the lepton sector. We explore experimental signatures of these
scenarios, which are observable in the future at ATLAS/CMS, LHCb, or Belle~II.

\end{abstract}

\maketitle

\section{Introduction}
\label{sec:intro}

Measurements of the $\bar{B} \to D\tau\bar{\nu}$ and $\bar{B} \to
D^*\tau\bar{\nu}$ decay rates are now available from BaBar~\cite{Lees:2012xj,
Lees:2013udz} and Belle~\cite{Huschle:2015rga} with their full datasets.
The $\bar{B} \to D^*\tau\bar{\nu}$ decay mode was also observed recently
by LHCb~\cite{Aaij:2015yra}.  These measurements are consistent with each other
and with earlier results~\cite{Aubert:2007dsa, Bozek:2010xy}, and together show
a significant deviation from Standard Model (SM) predictions for the combination
of the ratios
\begin{equation}
  \label{eq:Rdef}
  R(X) = \frac{\mathcal{B}(\bar{B} \to X \tau\bar{\nu})}
              {\mathcal{B}(\bar{B} \to X l\bar{\nu})}\, ,
\end{equation}
where $l = e, \mu$.  The measurements are consistent with $e/\mu$
universality~\cite{Aubert:2008yv, Dungel:2010uk}.  The $R(D^{(*)})$ data, their
averages~\cite{HFAGavg}, and the SM expectations~\cite{Lattice:2015rga,
Na:2015kha, Fajfer:2012vx} are summarized in Table~\ref{tab:data}\@.  (If the
likelihood of the measurements is Gaussian, then the deviation from the SM is
more than $4\sigma$.)  Kinematic distributions, namely the dilepton invariant
mass $q^2$, are also available from BaBar and Belle~\cite{Lees:2013udz,
Huschle:2015rga}, and must be accommodated by any model that modifies the rates.
In the future, Belle~II is expected to reduce the measured uncertainties
of $R(D^{(*)})$ by factors of $\sim\!5$ or more~\cite{Belle2predictions},
thereby driving experimental and theory precision to comparable levels.

\begingroup
\squeezetable
\begin{table}[b]
\begin{tabular}{c|ccc}
  \hline \hline
  &  $R(D)$  &  $R(D^*)$  & Corr.\\
  \hline
  BaBar                  & $0.440 \pm 0.058 \pm 0.042$
                         & $0.332 \pm 0.024 \pm 0.018$		&  $-0.27$\\
  Belle                  & $0.375 \pm 0.064 \pm 0.026$ 
                         & $0.293 \pm 0.038 \pm 0.015$		&  $-0.49$ \\
  LHCb                   &  & $0.336 \pm 0.027 \pm 0.030$  &  \\
  \hline
  Exp. average           & $0.391 \pm 0.041 \pm 0.028$ 
                         &  $0.322 \pm 0.018 \pm 0.012$     & $-0.29$\\ \hline
  SM expectation         &  $0.300 \pm 0.010$  &  $0.252 \pm 0.005$  \\          \hline
  Belle~II, 50/ab &  $\pm 0.010$        &  $\pm 0.005$\\
  \hline\hline
\end{tabular}
\caption{Measurements of $R(D^{(*)})$~\cite{Lees:2012xj, Huschle:2015rga,Aaij:2015yra},
         their averages~\cite{HFAGavg}, the SM predictions~\cite{Lattice:2015rga,
         Na:2015kha, Fajfer:2012vx}, and future sensitivity~\cite{Belle2predictions}.
         The first (second) experimental errors are statistical (systematic).}
\label{tab:data}
\end{table}
\endgroup

In the type-II two-Higgs-doublet model (2HDM), the $\bar{B} \to
D^{(*)}\tau\bar{\nu}$ rate (as well as $B^- \to \tau\bar{\nu}$) receives
contributions linear and quadratic
in $m_b\, m_\tau\tan^2\beta/m_{H^\pm}^2$~\cite{Krawczyk:1987zj,Kalinowski:1990ba,
Hou:1992sy}, which can be substantial if $\tan\beta$ is large.  However,
the $R(D^{(*)})$ data are inconsistent with this scenario~\cite{Lees:2012xj}.  

Discovering new physics (NP) in transitions between the third and second
generation fermion fields has long been considered plausible, since the flavor
constraints are weaker on four-fermion operators mediating such transitions.
(Prior studies of $B \to X_s\nu\bar{\nu}$~\cite{Grossman:1995gt} and $B_{(s)}
\to \tau^+\tau^-(X)$~\cite{Hewett:1995dk,Grossman:1996qj} decays were motivated
by this consideration.) However, $\bar{B} \to D^{(*)}\tau\bar{\nu}$ is mediated
by the tree-level $b \to c$ transition. It is suppressed in the SM neither
by CKM angles (compared to other $B$ decays) nor by loop factors, with only
a modest phase space suppression due to the $\tau$ mass. This goes against
the usual lore that the first manifestations of new physics at low energies
are most likely to occur in processes suppressed in the SM\@.

The goal of this paper is to explore flavor structures for NP capable of
accommodating the central values of the $R(D^{(*)})$ data summarized in
Table~\ref{tab:data}\@. To do so, a sizable NP contribution to semileptonic
$b \rightarrow c$ decays must be present, and the NP mass scale must be
near the weak scale. This requires nontrivial consistency with other
constraints, such as direct searches at the LHC and precision electroweak
data from LEP\@. When NP couplings to other generations are present, constraints
from flavor physics, such as meson mixing and rare decays, also play a role. For
example, any flavor model predicts some relation between the $\bar{b}c\,
\bar{\nu}\tau$ and $\bar{b}u\, \bar{\nu}\tau$ operators, so models explaining
$R(D^{(*)})$ must accommodate the observed $B^- \to \tau\bar{\nu}$ branching
ratio, which agrees with the SM~\cite{Lees:2012ju,Kronenbitter:2015kls}.  We
show below that despite strong constraints some scenarios remain viable and
predict signals in upcoming experiments.

We begin by presenting new inclusive calculations that demonstrate that the
measured central values of $R(D^{(*)})$ are in tension with the SM, independent
of form factor computations.  Then, in \sec{operator}, we perform a general
operator analysis to identify which four-fermion operators simultaneously fit
$R(D)$ and $R(D^*)$. In \sec{models} we discuss possible mediators that can
generate the viable operators. We identify working models with leptoquark
mediators that are minimally flavor violating in the quark sector, and we
confirm their consistency with current experimental constraints. Finally,
\sec{conc} contains our conclusions and a discussion of possible future signals
at the LHC and Belle~II\@.  Appendix~\ref{sec:U23} contains a discussion
of $U(2)^3$ models.

\subsection{Standard Model considerations}
\label{ssec:SMincl}

The tension between the central values of the $R(D^{(*)})$ data and the SM is
independent of the theoretical predictions for $R(D^{(*)})$ quoted in
Table~\ref{tab:data}\@.  The measured $R(D^{(*)})$ values imply a significant
enhancement of the inclusive $B \to X_c \tau \bar{\nu}$ rate, which can be
calculated precisely in the SM using an operator product expansion, with
theoretical uncertainties that are small and essentially independent from those
of the exclusive rates.

To see this, note that the isospin-constrained fit for the branching ratios
is quoted as~\cite{Lees:2012xj}
\beq\label{eq:sumoftwo}
  \mathcal{B}(\bar{B} \to D^*\tau\bar{\nu}) + 
  \mathcal{B}(\bar{B} \to D \tau\bar{\nu}) = (2.78 \pm 0.25) \%\,,
\eeq
which applies for $B^\pm$ decays (recall the lifetime difference of $B^\pm$ and
$B^0$).  The averages in Table~\ref{tab:data} imply for the same quantity the
fully consistent result,
\beq\label{eq:sumoftwoB}
  \mathcal{B}(\bar{B} \to D^*\tau\bar{\nu}) + 
  \mathcal{B}(\bar{B} \to D \tau\bar{\nu}) = (2.71 \pm 0.18) \%\,.
\eeq

The SM prediction for $R(X_c)$, the ratio for inclusive decay rates, can
be computed in an operator product expansion.  Updating results in
Refs.~\cite{Falk:1994gw, Hoang:1998hm}, and including the two-loop QCD
correction~\cite{Biswas:2009rb}, we find
\beq\label{Rinclusive}
  R(X_c) = 0.223 \pm 0.004\,.
\eeq
The uncertainty mainly comes from $m_b^{1S}$, the HQET matrix element
$\lambda_1$, and assigning an uncertainty equal to half of the order
$\alpha_s^2$ term in the perturbation series in the $1S$
scheme~\cite{Bauer:2004ve}. The most recent world average, $\mathcal{B}(B^-
\to X_c e \bar{\nu}) = (10.92 \pm 0.16)\%$~\cite{Bernlochner:2012bc,
Amhis:2014hma}, then yields the SM prediction,
\beq\label{inclprediction}
  \mathcal{B}(B^- \to X_c\tau\bar{\nu}) = (2.42 \pm 0.05)\%\,.
\eeq

In $B^-\to X_c e \bar{\nu}$ decay, hadronic final states other than $D$ and
$D^*$ contribute about 3\% to the 10.92\% branching ratio quoted above, and
the four lightest orbitally excited $D$ meson states (often called collectively
$D^{**}$) account for about 1.7\%.  Using Ref.~\cite{Leibovich:1997em} for the
theoretical description of these decays, taking into account the phase space
differences and varying the relevant Isgur-Wise functions, suggests
$R(D^{**}) \gtrsim 0.15$ for the sum of these four states.  This in turn implies
for the sum of the central values of the rates to the six lightest charm meson
states
\beq\label{sixmodes}
\mathcal{B}(\bar{B} \to D^{(*)}\tau\bar\nu) 
  + \mathcal{B}(\bar{B} \to D^{**}\tau\bar\nu) \sim3\%\,,
\eeq
in nearly $3\sigma$ tension with the inclusive calculation in
Eq.~(\ref{inclprediction}). Note that Eqs.~(\ref{eq:sumoftwo}),
(\ref{eq:sumoftwoB}), and (\ref{sixmodes}) are also in mild tension with the LEP
average of the rate of an admixture of $b$-flavored hadrons to decay to $\tau$
leptons~\cite{PDG},
\beq
  \mathcal{B}(b \to X\tau^+\nu) = (2.41 \pm 0.23)\%\,.
\eeq
Since both the experimental and theoretical uncertainties of  $\mathcal{B}(B \to
X_c\tau\bar\nu)$ are different from the exclusive rates, its direct measurement
from Belle and BaBar data would be interesting and timely~\cite{Ligeti:2014kia}.

\section{\boldmath $\bar{B}\to D^{(*)}\tau\bar\nu$ operator analysis}
\label{sec:operator}

\begin{table*}[tb]
\begin{tabular}{l|ccc|cc}
  \hline\hline
	& Operator & & Fierz identity & 
    Allowed Current & $\delta\mathcal{L}_\text{int}$ \\[2pt]
  \hline
  $\mathcal{O}_{V_L}$   & $(\bar{c} \gamma_\mu P_L b)\,(\bar{\tau} \gamma^\mu P_L \nu)$ & & &
    $(\repone,\repthr)_0$  &
    ($g_q\bar{q}_L {\bm\tau} \gamma^\mu q_L + g_\ell\bar{\ell}_L {\bm\tau} \gamma^\mu\ell_L) W'_{\mu} $\\
  $\mathcal{O}_{V_R}$   & $(\bar{c} \gamma_\mu P_R b)\,(\bar{\tau} \gamma^\mu P_L \nu)$ & & & & \\
  $\mathcal{O}_{S_R}$   & $(\bar{c} P_R b)\,(\bar{\tau} P_L \nu)$ & & &\\
  $\mathcal{O}_{S_L}$   & $(\bar{c} P_L b)\,(\bar{\tau} P_L \nu)$ & & &
    \multirow{-2}*{$\bigg\rangle(\repone,\reptwo)_{1/2}$} &
    \multirow{-2}*{$(\lambda_d\bar{q}_L d_R \phi + \lambda_u\bar{q}_L u_R i\tau_2 \phi^\dag
                     + \lambda_\ell\bar{\ell}_L e_R \phi)$}\\
  $\mathcal{O}_T$       & $(\bar{c}\sigma^{\mu\nu}P_L b)\,(\bar{\tau}\sigma_{\mu\nu}P_L \nu)$ & & & & \\[2pt]
  \hline
  \multirow{2}*{$\mathcal{O}'_{V_L}$} &
    \multirow{2}*{$(\bar{\tau} \gamma_\mu P_L b)\,(\bar{c} \gamma^\mu P_L \nu)$} &
    \multirow{2}*{$\longleftrightarrow$} & \multirow{2}*{$\mathcal{O}_{V_L}\bigg\langle$} &
    $(\repthr,\repthr)_{2/3}$   & $\lambda\,\bar{q}_L {\bm\tau} \gamma_\mu \ell_L {\bm U}^\mu $\\
  & & & &\\
  $\mathcal{O}'_{V_R}$  & $(\bar{\tau} \gamma_\mu P_R b)\,(\bar{c} \gamma^\mu P_L \nu)$ &
    $\longleftrightarrow$ & $-2\mathcal{O}_{S_R}$ &
    \multirow{-2}*{$\bigg\rangle(\repthr,\repone)_{2/3}$} &
    \multirow{-2}*{$(\lambda\,\bar{q}_L \gamma_\mu \ell_L
                     + \tilde{\lambda}\,\bar{d}_R \gamma_\mu e_R) U^\mu$}\\
  $\mathcal{O}'_{S_R}$  & $(\bar{\tau} P_R b)\,(\bar{c} P_L \nu)$ &
    $\longleftrightarrow$ & $-\frac{1}{2}\mathcal{O}_{V_R}$ & \\
  $\mathcal{O}'_{S_L}$  & $(\bar{\tau} P_L b)\,(\bar{c} P_L \nu)$ &
    $\longleftrightarrow$ & $-\frac{1}{2}\mathcal{O}_{S_L} - \frac{1}{8}\mathcal{O}_T$ &
    $(\repthr,\reptwo)_{7/6}$ &
    $(\lambda\,\bar{u}_R \ell_L + \tilde{\lambda}\,\bar{q}_L i\tau_2 e_R) R $\\
  $\mathcal{O}'_T$      & $(\bar{\tau}\sigma^{\mu\nu}P_L b)\,(\bar{c}\sigma_{\mu\nu}P_L \nu)$ &
    $\longleftrightarrow$ & $-6\mathcal{O}_{S_L} + \frac{1}{2}\mathcal{O}_T$ & & \\[2pt]
  \hline
  $\mathcal{O}''_{V_L}$ & $(\bar{\tau} \gamma_\mu P_L c^c)\,(\bar{b}^c \gamma^\mu P_L \nu)$ &
    $\longleftrightarrow$ & $-\mathcal{O}_{V_R}$ & & \\
  $\mathcal{O}''_{V_R}$ & $(\bar{\tau} \gamma_\mu P_R c^c)\,(\bar{b}^c \gamma^\mu P_L \nu)$ &
    $\longleftrightarrow$ & $-2\mathcal{O}_{S_R}$ &
    $(\arepthr,\reptwo)_{5/3}$ &
    $(\lambda\,\bar{d}_R^c\gamma_\mu\ell_L+\tilde{\lambda}\,\bar{q}_L^c\gamma_\mu e_R)V^\mu$\\
  \multirow{2}*{$\mathcal{O}''_{S_R}$} &
    \multirow{2}*{$(\bar{\tau} P_R c^c)\,(\bar{b}^c P_L \nu)$} &
    \multirow{2}*{$\longleftrightarrow$} & \multirow{2}*{$\frac{1}{2}\mathcal{O}_{V_L}\bigg\langle$} &
    $(\arepthr,\repthr)_{1/3}$ & 
    $\lambda\, \bar{q}_L^c i \tau_2 {\bm\tau} \ell_L {\bm S} $\\
  & & & &\\
  $\mathcal{O}''_{S_L}$ & $(\bar{\tau} P_L c^c)\,(\bar{b}^c P_L \nu)$ &
    $\longleftrightarrow$ & $-\frac{1}{2}\mathcal{O}_{S_L} + \frac{1}{8}\mathcal{O}_T$ &
    \multirow{-2}*{$\bigg\rangle(\arepthr,\repone)_{1/3}$} &
    \multirow{-2}*{$(\lambda\,\bar{q}_L^c i\tau_2\ell_L+\tilde{\lambda}\,\bar{u}_R^c e_R) S$}\\
  $\mathcal{O}''_T$     & $(\bar{\tau}\sigma^{\mu\nu}P_L c^c)\,(\bar{b}^c\sigma_{\mu\nu}P_L \nu)$ &
    $\longleftrightarrow$ & $-6\mathcal{O}_{S_L} - \frac{1}{2}\mathcal{O}_T$ & & \\[2pt]
  \hline\hline
\end{tabular}
\caption{All possible four-fermion operators that can contribute to $\bar B \to
D^{(*)} \tau\bar{\nu}$. Operators for which no quantum numbers are given can
only arise from dimension-8 operators in a gauge invariant completion. For other
operators the interaction terms which are subsequently integrated out are given.
For the $T$ operators we use the conventional definition of $\sigma^{\mu\nu} =
i[\gamma^\mu,\gamma^\nu]/2$.}
\label{tab:ops}
\end{table*}

In this section we study operators mediating $b \to c\tau\bar{\nu}$
transitions. In contrast to prior operator fits~\cite{Fajfer:2012jt,
Tanaka:2012nw,Sakaki:2013bfa,Biancofiore:2013ki}, we adopt an overcomplete
set of operators corresponding to all possible contractions of spinor
indices and Lorentz structures to help with the classification of viable
models. (We also take into account the constraints from $q^2$ spectra, which
were unavailable at the time of the first operator analyses.) Although Fierz
identities allow different spinor contractions to be written as linear
combinations of operators with one preferred spinor ordering, the set of
possible currents that can generate the operators is manifest
in the overcomplete basis.

We parametrize the NP contributions by
\begin{equation}
\mathcal{H} = \frac{4G_F}{\sqrt{2}}\, V_{cb}\, \mathcal{O}_{V_L}
	+ \frac{1}{\Lambda^2} \sum_{i}
	C^{(\prime,\prime\prime)}_i\, \mathcal{O}^{(\prime,\prime\prime)}_i \,.
\end{equation}
(Throughout this paper we do not display Hermitian conjugates added
to interaction terms as appropriate.) Here the primes denote different ways
of contracting the spinors, as shown in \tab{ops}, which also presents their
Fierz transformed equivalents in terms of the ``canonically'' ordered fields
(unprimed operators). In the SM, only the $\mathcal{O}_{V_L}$ operator is
present. (For illustration, the type-II 2HDM generates the operator
$\mathcal{O}_{S_R}$ with $C_{S_R}/\Lambda^2 = -2\sqrt2\, G_F V_{cb}\,
m_b\, m_\tau\tan^2\beta/m_{H^\pm}^2$.)

We do not consider the possibility of the neutrino being replaced by  another
neutral particle, such as a sterile neutrino, which yields
additional operators. The large enhancement of an unsuppressed SM rate favors NP
that can interfere with the SM\@. A non-SM field in the final state would preclude
the possibility of interference, leading to larger Wilson coefficients and/or
lower mass scales for the NP, making the interpretation in terms of concrete
models more challenging.

We assume that the effects of NP can be described by higher dimension operators
respecting the SM gauge symmetries. This is only evaded if the NP mediating
these transitions is light or if it is strongly coupled at the electroweak
scale; in either case there are severe constraints. We classify operators by the
representations under $SU(3)_C \times SU(2)_L \times U(1)_Y$ of the mediators
that are integrated out to generate them, as shown in the last column of
\tab{ops}\@. Some mediators uniquely specify a single operator, while others can
generate two simultaneously.  Anticipating the large Wilson coefficients
necessary to fit the observed $R(D^{(*)})$ ratios, we focus on operators which
can arise from dimension-6 gauge-invariant terms.  Operators which can only come
from SM gauge invariant dimension-8 terms or cannot be generated by integrating
out a low-spin mediator will be omitted. Such contributions would be suppressed
by additional powers of $v/\Lambda$, or could only arise from strongly coupled
NP\@.

We calculate the contributions of all operators in the heavy quark
limit~\cite{Isgur:1989vq}. Subleading corrections Our method follows that
of Ref.~\cite{Goldberger:1999yh}, and we rederived and confirmed those results.
(A missing factor of $(1-m_\ell^2/q^2)$ has to be inserted in Eq.~(10) of
Ref.~\cite{Goldberger:1999yh}.)  We use the most precise single measurement of
the $h_{A_1}$ form factor~\cite{Dungel:2010uk}, which equals the Isgur--Wise
function in the heavy quark limit. 

Higher order corrections are neglected, except for the following two effects
that are known to be significant.  For scalar operators a numerically sizable
term, $(m_B+m_{D^*})/(m_b+m_c) \simeq 1.4$, arises from $\bra{D^*} \bar{c}
\gamma_5 b \ket{B} = -q^\mu \bra{D^*} \bar{c} \gamma_\mu\gamma_5 b \ket{B} /
(m_b+m_c)$. We also include the leading-log scale dependence of scalar (here
$C_S$ is either $C_{S_L}$ or $C_{S_R}$) and tensor
currents~\cite{Dorsner:2013tla} in fits to models where they appear
simultaneously,
\begin{align}
\label{eq:CSTrunning}
  C_S(m_b) &= \left( \frac{\alpha_s(m_t)}{\alpha_s(m_b)} \right)^{-12/23}
	\left( \frac{\alpha_s(M)}{\alpha_s(m_t)} \right)^{-12/21}
	C_S(M)\,, \nn\\
  C_T(m_b) &= \left( \frac{\alpha_s(m_t)}{\alpha_s(m_b)} \right)^{4/23}
	\left( \frac{\alpha_s(M)}{\alpha_s(m_t)} \right)^{4/21}
	C_{T}(M)\,.
\end{align}
For numerical calculations we use a reference scale $M = 750 \GeV$\@.  The
sensitivity to this choice is small, as most of the running occurs at low
scales, between $m_b$ and $m_t$.

To test the robustness of our results to $\mathcal{O}(\lqcd/m_{c,b})$
corrections, we varied the slope parameter of the Isgur--Wise function, $\rho^2$,
by $\pm 0.2$ (motivated by Ref.~\cite{Grinstein:2001yg}), and found less than
$1\sigma$ change in the results.  We leave consideration of  ${\cal
O}(\lqcd/m_{c,b})$ corrections for the new physics, of the sort carried out
for 2HDMs in \cite{Grossman:1994ax,Grossman:1995yp}, for future work.

\begin{figure*}[tb]
\includegraphics[width=.95\textwidth]{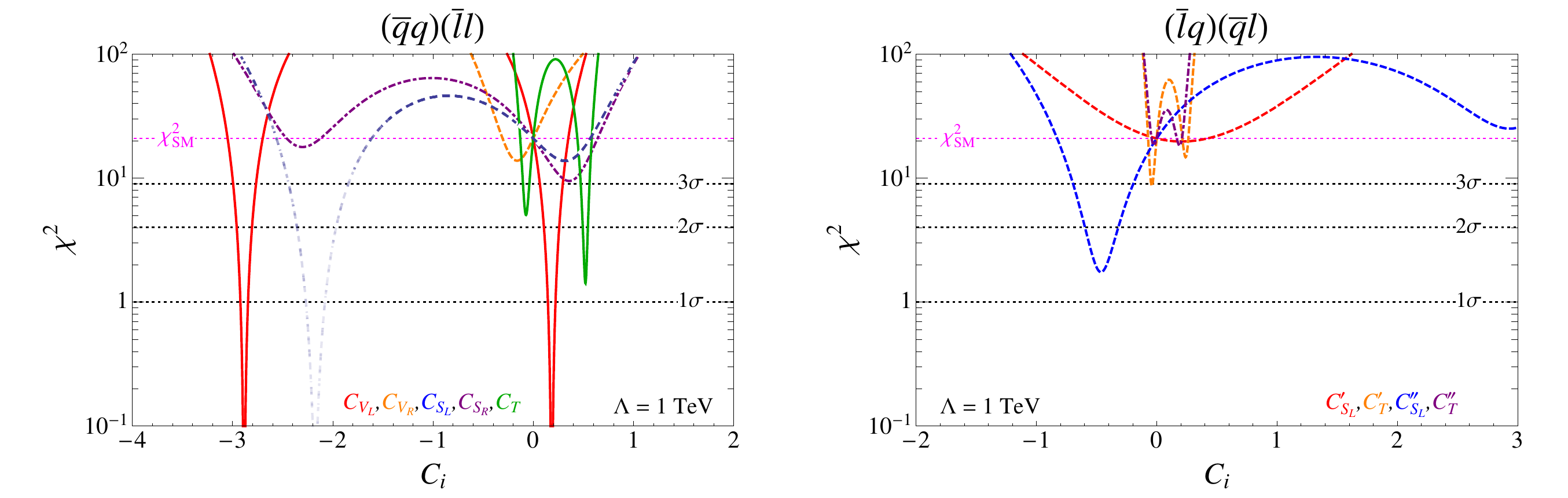}
\caption{Goodness-of-fit for the coefficients of individual operators from the
measured $R(D)$ and $R(D^*)$ ratios.  Besides the fits to the unprimed operators
in Table~\ref{tab:ops} (left), we also show fits to primed operators not related
by simple rescalings (right). Faded regions for $C_{S_L}$ indicate good fits to
the observed rates excluded by the measurement of the $q^2$
spectrum~\cite{Lees:2013udz}.
Note that the $\chi^2$ includes experimental and SM theory uncertainties, but
not theory uncertainties on NP\@.}
\label{fig:1Dfit}
\end{figure*}

\begin{figure*}[tb]
\includegraphics[width=.95\textwidth]{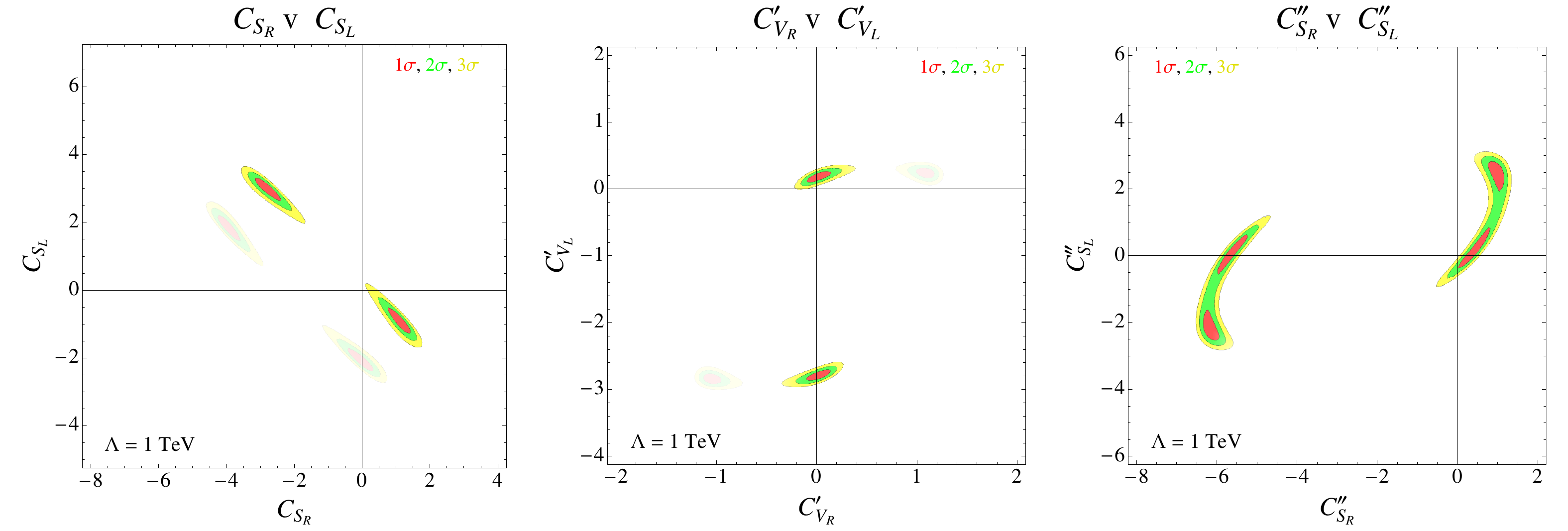}
\caption{Goodness-of-fit for coefficients of operators which can be generated
from dimension-6 operators with fermion bilinears having the same SM quantum
numbers.  The plots show 1-, 2-, and 3$\sigma$ allowed regions.  Approximate
regions of parameter space excluded by the measurement of the $q^2$
spectrum~\cite{Lees:2013udz} are presented as faded regions, as in \fig{1Dfit}.}
\label{fig:2Dfit}
\end{figure*}

Figure~\ref{fig:1Dfit} shows the results of $\chi^2$ fits to $R(D)$ and
$R(D^*)$ for each of the four-fermion operators in \tab{ops} individually. 
Here and below, our $\chi^2$ includes experimental and SM theory uncertainties,
but does not include theory uncertainties on NP, which are subdominant.
Throughout, we assume that no new large sources of CP violation are present, i.e., we assume that the
the phases of the NP operators are aligned with the phase of the SM vector operator. A contribution
from the $\mathcal{O}_T$ operator or a modification of the SM contribution
proportional to $\mathcal{O}_{V_L}$ (or any of its equivalents under Fierz
identities) provide good fits to the data.  The $\mathcal{O}_{S_L}$ operator
can also fit the total rates, but it leads to $q^2$ spectra
incompatible with observations.  The operator $\mathcal{O}''_{S_L}$ also gives
a good fit, which is not apparent from only considering the unprimed operators.
(Note that the measurements of $R(D)$ and $R(D^*)$ depend
on the operator coefficients, because the decay distributions are modified
by the new physics contribution, affecting the experimental efficiencies and
the measured rates~\cite{Lees:2012xj,Manuel}. This effect cannot be included
in our fits, providing another reason to take the $\chi^2$ values as rough
guides only.)

As noted earlier, certain mediators can generate two contributing operators
simultaneously. \fig{2Dfit} shows the three such two-dimensional $\chi^2$ fits.
While any two rates can be explained by fitting two operator coefficients, the
existence of a solution consistent with all other constraints with a given
flavor structure is nontrivial and is the topic of the following section.
A summary of all coefficients of best fit points with $\chi_\text{min}^2 < 5$ and
acceptable $q^2$ spectra is provided in \tab{bestfit}. 

\begin{table}[bt]
\tabcolsep 6pt
\begin{tabular}{c|l}
\hline\hline
Coefficient(s)  &  Best fit value(s) ($\Lambda=1\TeV$)  \\
\hline
$C_{V_L}$  &  $0.18 \pm 0.04$, \quad $-2.88 \pm 0.04$  \\
$C_T$  &  $0.52 \pm 0.02$, \quad $-0.07 \pm 0.02$    \\
$C''_{S_L}$  &  $-0.46 \pm 0.09$   \\
\hline
$(C_R,\, C_L)$  &  $(1.25,-1.02)$, \quad $(-2.84,3.08)$ \\
$(C'_{V_R},\, C'_{V_L})$  &  $(-0.01,0.18)$, \quad $(0.01,-2.88)$ \\
$(C''_{S_R},\, C''_{S_L})$  &  $(0.35,-0.03)$, \quad $(0.96,2.41)$, \\
   &  $(-5.74, 0.03)$, \quad $(-6.34,-2.39)$ \\
\hline\hline
\end{tabular}
\caption{Best-fit operator coefficients with acceptable $q^2$~spectra
and $\chi_\text{min}^2 < 5$. For the 1D fits in Fig.~\ref{fig:1Dfit}
we include the $\Delta\chi^2 < 1$ ranges (upper part), and show
the central values of the 2D fits in Fig.~\ref{fig:2Dfit} (lower part).}
\label{tab:bestfit}
\end{table}

\begin{figure}[btp]
\includegraphics[width=.95\columnwidth]{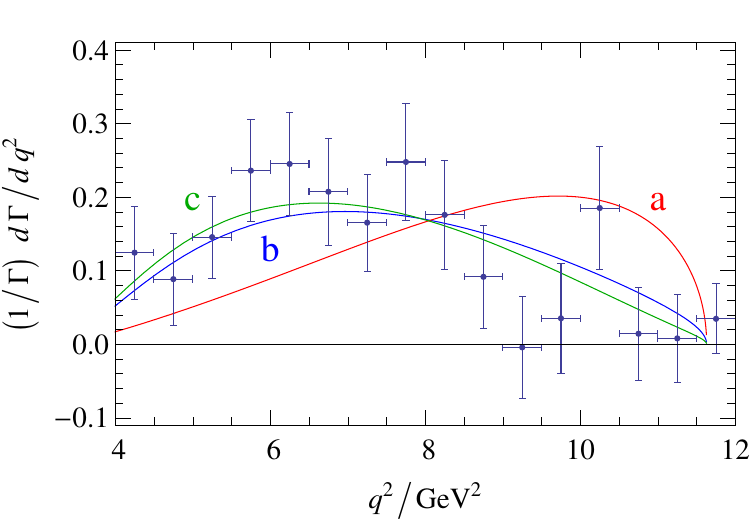}
\caption{Normalized $\d\Gamma(\bar B\to D\tau\bar\nu)/\d q^2$ distributions. 
The histogram shows the BaBar data~\cite{Lees:2013udz}. The red, blue, and green
curves correspond to models with Wilson coefficients in the (a) top, (b) middle,
and (c) bottom lines in \tab{mfv}\@.}
\label{fig:q2plot}
\end{figure}

Besides the branching ratios, additional model discrimination comes from the
$q^2$ spectra (especially in $\bar B\to D\tau\bar\nu$), which are consistent
with SM expectations~\cite{Lees:2013udz,Huschle:2015rga}. It is not possible
to do a combined fit with publicly available data, because correlations among
different $q^2$ bins are unavailable. We follow Ref.~\cite{Lees:2013udz}
in eliminating certain models by comparing their predicted $q^2$ spectra
with the measurement. It was observed that two of the four solutions
in the $C_{S_R}$--\,$C_{S_L}$ plane (\fig{2Dfit}, left plot) are
excluded~\cite{Lees:2013udz}, as indicated by the faded regions.
In the $C'_{V_R}$--\,$C'_{V_L}$ plane (middle plot), we find the measured $q^2$
spectra exclude regions that provide good fits to the total rates for values
of $|C'_{V_R}| \gtrsim 0.5$. In the $C''_{S_R}$--\,$C''_{S_L}$ plane (right
plot) all fits consistent with the total rates are also consistent
with the $q^2$ spectra.  Figure~\ref{fig:q2plot} illustrates this by showing
the measured $\d\Gamma(B \to D\tau\bar{\nu})/ \d q^2$ spectrum together
with one disfavored and two viable models, corresponding to the entries
in \tab{mfv}\@.

\begin{table}[btp]
\tabcolsep 6pt
\begin{tabular}{c|llc}
\hline\hline
Label  &  \multicolumn{2}{c}{Coefficients ($\Lambda=1\TeV$)}  &  Comment  \\
\hline
(a) & $C'_{V_R} = 1.10$ & $C'_{V_L} = 0.24$     &  disfavored  \\
(b)&$C'_{V_R} = -0.01$ & $C'_{V_L} = 0.18$    &  allowed \\
(c)&$C''_{S_R} = 0.96$  &  $C''_{S_L} =2.41$  &  allowed \\
\hline\hline
\end{tabular}
\caption{Operator coefficients for the $q^2$ spectra in Fig.~\ref{fig:q2plot}\@.}
\label{tab:mfv}
\end{table}

\section{Models}
\label{sec:models}

Having identified dimension-6 operators that can generate the observed $\bar{B}
\to D^{(*)}\tau\bar{\nu}$ rate, we now consider possible UV completions. The
TeV-scale required for the four-fermion operators points to a tree-level NP
contribution. Concerning the flavor structure, one possibility is that NP is
aligned with the SM Yukawa matrices and in the fermion mass basis only gives
rise to the $\bar{b}c\, \bar{\nu}\tau$ four-fermion interaction.  In such a
scenario the measurement of $R(D^{(*)})$ would have little interplay with other
flavor data; however, it is hard to imagine a UV completion with such precise
alignment with the SM flavor structure.  On the other hand, if operators are
generated with general fermion content, $P_{ijkl}\, \bar{d}_i u_j\, \bar{\nu}_k
\ell_l$, where $i,j,k,l$ are generation indices, then some $P_{ijkl}$
coefficients must satisfy strong experimental constraints. Most of the
completions we consider below have also been studied, but without considerations
to their possible flavor structure; see, e.g., Refs.~\cite{Fajfer:2012jt,
Tanaka:2010se, Tanaka:2012nw, Datta:2012qk, Sakaki:2012ft, Sakaki:2013bfa,
Dorsner:2011ai, Dorsner:2011ai, Kamenik:2008tj}

Independent of the UV completion, flavor-anarchic couplings are excluded by
other rare decays and CKM unitarity constraints~\cite{Carpentier:2010ue}. We
therefore consider NP that is charged under a flavor symmetry, focusing on the
possibility of Minimal Flavor Violation (MFV)~\cite{Chivukula:1987py,
Hall:1990ac, D'Ambrosio:2002ex}.  In the quark sector, the MFV framework assumes that the
breaking of the $U(3)_Q \times U(3)_u \times U(3)_d$ flavor symmetry has a
single source in both the SM and beyond. It is parametrized by the Yukawas,
acting as spurions of the symmetry breaking, transforming in the
$(\repthr,\arepthr,\repone)$ representation for $Y_u$ and the
$(\repthr,\repone,\arepthr)$ for $Y_d$. Focusing primarily on MFV in the quark
sector, for viable models we attempt to extend the MFV scenario to the lepton
sector as well. We find a unique model that is consistent with MFV both in the
quark and the lepton sector, and comment on alternative approaches involving
horizontal symmetries. We begin by showing in \ssec{uncolor} that uncolored
mediators are disfavored by other constraints.  Then, in \ssec{leptoquark}, we
identify viable MFV models with leptoquark mediators.

\subsection{Uncolored mediators}
\label{ssec:uncolor}

\subsubsection{Higgs-like scalars}
\label{ssec:newscalar}

We first explain why no new color-neutral scalar, such as those in the
nonstandard 2HDMs proposed
in Refs.~\cite{Crivellin:2012ye,Celis:2012dk,Ko:2012sv}, is compatible
with an MFV structure. The simplest case, a flavor singlet scalar, would be
constrained to have couplings proportional to the SM Higgs. As shown
in \fig{2Dfit}, comparable values of $C_{S_L}$ and $C_{S_R}$ are needed to explain
the current $\bar B \to D^{(*)} \tau\bar{\nu}$ data.  $C_{S_L}$ is proportional to $y_c$, which implies that $\bar B \to D^{(*)} \tau\bar{\nu}$ cannot be fit, keeping the charged Higgs heavier than collider limits, consistently with perturbativity of $y_t$.

An alternative is to charge the scalar under the quark flavor symmetries. In
this case, the scalar needs to be in the representation of some combination of
Yukawa spurions, since it must couple to leptons as well.  To generate both
scalar couplings simultaneously, an unsuppressed $\mathcal{O}(1)$ coupling to
one chiral combination of first-generation quarks is unavoidable.   Integrating out the scalar generates dangerous 
four-quark and two-quark two-lepton operators. Four-quark operators of this
type are strongly constrained, and such a scalar would also appear in $\tau^+ \tau^-$ resonance searches~\cite{Aad:2015osa,CMS:2015ufa}.

\subsubsection{$W'$-like vector triplets}
\label{ssec:Wprime}

A rescaling of the SM operator, $\mathcal{O}_{V_L}$, provides a good fit to the
data. A simple possibility is then the presence of a $W'$, a new vector that
couples similarly to the SM $W$ boson. To avoid explicitly breaking the SM gauge
symmetry, a full $SU(2)_L$ triplet is needed. The simplest choice, a flavor
singlet $W'$, is tightly constrained by the LHC~\cite{Chatrchyan:2012gqa,
Aad:2014xra, Aad:2014xea}, with electroweak-strength coupling of a $W'$ to first
generation quarks excluded up to $m_{W'} \sim 1.8\TeV$. This is in conflict with
the $0.2 \sim g^2\, |V_{cb}|\, (1\TeV/m_{W'})^2$ coupling needed to explain the
$\bar B\to D^{(*)}\tau\bar\nu$ data.

As with the scalars, it is possible to consider a $W'$ that transforms
under a nontrivial representation of the $[U(3)]^3$ quark flavor symmetry
group, in an attempt to suppress the couplings to light quark generations.
However, the $W'$ needs to simultaneously couple to leptons, fixing its
flavor assignment to be the conjugate of some combination of Yukawa spurions.
This means that the same flavor structure allowing the $W'$ to couple
to leptons can always couple to a current of left-handed quarks contracted
as a flavor singlet, allowing for universal couplings to all quark generations.

Suppose that the flavor singlet universal coupling to quarks is accidentally
small. The minimal charge assignments are $(\arepthr,\repthr,\repone)$ and
$(\arepthr,\repone,\repthr)$. In the $(\arepthr,\repthr,\repone)$ case,
(semi)leptonic $b \to c$ transitions are suppressed by  $y_c$, and therefore
the central values of $R(D^{(*)})$ cannot be generated with perturbative
couplings.  In the ($\arepthr,\repone,\repthr)$ case, the $R(D^{(*)})$ data
can be accommodated by the interaction $\delta\mathcal{L} =
g\,\bar{q}_L^i Y_d^{ij} \bm{\tau \cdot W}_\mu^{jk} \gamma^\mu q_L^k$ and
an equivalent coupling to the leptons with the flavor indices of $Y_d$ and
$\bm{W}_\mu$ contracted into each other, given $y_b = {\cal O}(1)$. LHC direct production bounds are
then evaded.  However, tree-level FCNCs are mediated by the
neutral component, $W^0$, and fitting the $\bar{B} \to D^{(*)}\tau\bar{\nu}$
rates implies too large of a contribution to $B$--$\bar{B}$ mixing.
For an approach using a combination of dynamical singlet suppression and reduced
symmetries to evade these issues, see Ref.~\cite{Greljo:2015mma}.

\subsection{Theories of MFV leptoquarks}
\label{ssec:leptoquark}

The second class of potential tree-level mediators are leptoquarks. With no way
to contract a single quark representation with Yukawa spurions to form a flavor
singlet, leptoquarks in the MFV framework must carry some charges
under the quark flavor symmetries for flavor indices to be contracted
in interaction terms. Additionally, for representations of the flavor group
to be coupled together in a way that gives non-zero couplings, the mediators
must come in fundamental representations of some of the $[U(3)]^3$ quark flavor
group. Any larger representation will always either give vanishing coupling when
contracted with Yukawa spurions in order to couple to quarks or have the spurions
entirely contracted into the leptoquark so as to not generate any new couplings.
For reviews of leptoquarks, see, e.g., Refs.~\cite{Buchmuller:1986zs,
Davidson:1993qk}, and for discussion of MFV leptoquarks, see, e.g.,
Ref.~\cite{Davidson:2010uu}.

The six simplest possible flavor assignments under 
$U(3)_Q \times U(3)_u \times U(3)_d$, for scalar leptoquarks, are
\begin{equation}
  S \sim (\arepthr,\repone,\repone)\,, \quad
  S \sim (\repone,\arepthr,\repone)\,, \quad
  S \sim (\repone,\repone,\arepthr)\,,
\end{equation}
while vector leptoquarks can have charge
\begin{equation}
  U_\mu \sim (\repthr,\repone,\repone)\,, \quad
  U_\mu \sim (\repone,\repthr,\repone)\,, \quad
  U_\mu \sim (\repone,\repone,\repthr)\,.
\end{equation}
For each of these six charge assignments, the leptoquarks could be either
electroweak $SU(2)_L$ singlets or triplets.  Unless otherwise specified, all
comments below apply for both cases.  Since these leptoquarks need to mediate
processes involving different generations of quarks, additional experimental
constraints are important. We will see that the dominant bounds come
from the tension between allowing for a large enough effect to fit
the $R(D^{(*)})$ data, while being consistent with precision electroweak
constraints and contributions to FCNC decays, such as $\bar{B} \to
X_s \nu\bar{\nu}$.

Most choices of representation under the flavor groups can be immediately
discarded, since they either only generate Yukawa-suppressed operators for the
required $b \to c\tau\bar{\nu}$ decays, or have unsuppressed couplings to
first-generation quarks. The latter case would lead to new contributions to
$\tau^+\tau^-$ production at the LHC, which is severely constrained by $Z'$
searches~\cite{Aad:2015osa,CMS:2015ufa}. We estimate that couplings larger than
$\lambda/m \sim 0.25/\text{TeV}$ are excluded due to production of
$\tau^+\tau^-$ through $t$-channel leptoquark exchange. Below, we only consider
models that satisfy both of these constraints.

\subsubsection{Scalar leptoquarks --- $\tau$ alignment}
\label{sec:sLQ}

Consider a scalar leptoquark, $S$, with $U(3)_Q \times U(3)_u \times U(3)_d$
charge $(\bm{1},\bm{1},\bm{\bar{3}})$. The interaction terms with the smallest
number of Yukawa insertions for the electroweak singlet case take the form
\begin{align}
\label{eq:S113}
  \delta\mathcal{L}
    &= S(\lambda Y_d^\dagger\, \bar{q}^c_{L} i\tau_2 \ell^{\phantom c}_L
       + \tilde{\lambda} Y_d^\dagger Y_u\,\bar{u}^c_{R} e^{\phantom c}_R) \\
    &= S_i (\lambda y_{d_i} V^*_{ji} \,\bar{u}^c_{Lj} \tau^{\phantom c}_L
     - \lambda y_{d_i} \bar{d}^c_{Li} \nu^{\phantom c}_L
     + \tilde{\lambda} y_{d_i} y_{u_j} V^*_{ji}\, \bar{u}^c_{Rj} \tau^{\phantom c}_R)\,. \nn
\end{align}
Couplings to light quarks are suppressed by quark masses and/or small CKM angles,
so LHC bounds involving production off valence quarks are evaded.  Meanwhile,
$b \to c$ operators receive contributions of the form
\begin{equation}
\label{eq:S113Bop}
  \frac{C''_{S_R}}{\Lambda^2} = \frac{V^*_{cb}}{m_{S_3}^2} \lambda^2 y_b^2\,, \qquad
  \frac{C''_{S_L}}{\Lambda^2} = \frac{V^*_{cb}}{m_{S_3}^2} \lambda\tilde{\lambda}\,y_c y_b^2\,.
\end{equation}
This contribution can be sizable if $y_b = {\cal O}(1)$, which can follow
from a type II 2HDM at large $\tan \beta$. Due to the large suppression
of $C''_{S_L}$ from $y_c$, we use from Fig.~\ref{fig:2Dfit}, $C''_{S_L} \approx
0$ and $C''_{S_R}/\Lambda^2 = (0.35 \pm 0.1) /(1\TeV)^2$, corresponding
to $\lambda y_b/m_{S_3} \approx (2.95 \pm 0.6)/\text{TeV}$,
with $\tilde{\lambda}$ arbitrary as long as it remains perturbative.
The analogous term for the electroweak triplet scalar only generates
the $C''_{S_R}$ term, but with the opposite sign, due to the presence
of an extra $\tau^3$ factor in the interaction Lagrangian for the relevant
component. We do not study the triplet further, since under the assumption
of MFV generating an opposite sign contribution for $R(D^{(*)})$ is impossible.
(In this paper we neglect the possibility of complex couplings otherwise $CP$ violation constraints need to be studied,
which go beyond the scope of this paper and are unlikely to result in qualitatively
different allowed scenarios.) Similar operators contribute to $B^- \to \tau\bar{\nu}$
decay, with $V_{cb}$ replaced by $V_{ub}$. The contribution of these operators
to the measured rate is consistent with current data.

Coupling exclusively to third generation leptons can be enforced by imposing
a horizontal flavor symmetry, $U(1)_\tau$, under which the leptoquark and
third-generation lepton fields are equally and oppositely charged. Neutrino
masses and mixings require additional spurions that break the lepton flavor
symmetry, but these effects can be small enough to ignore in the present
discussion. For an example where such symmetries generate the entire structure
of $b \to c \tau\bar{\nu}$ transitions, see Ref.~\cite{Deshpande:2012rr},
and Refs.~\cite{Bhattacharya:2014wla,Calibbi:2015kma} for discussions
of lepton symmetries in $b \to c$ transitions in relation to other anomalies.

The presence of a large $S_3 \bar{t}^c \tau$ coupling also generates a deviation
from the SM prediction for the $Z\tau^+\tau^-$ coupling, which can affect both
the total rate and asymmetry of $Z \to \tau^+ \tau^-$ decays. For scalar
leptoquarks, such corrections are finite and calculable~\cite{Mizukoshi:1994zy}.
In the case where only the couplings proportional to $\lambda$ are present,
these provide a bound on $\lambda y_b/m_{S_3}$ comparable to the value needed
to fit the $R(D^{(*)})$ data. However, interference between the $\lambda$ and
$\tilde{\lambda}$ couplings can be large, since the latter is not suppressed
by $y_c$ (unlike in the decay of the $b$-quark). If $|\tilde{\lambda}/\lambda|
\gtrsim 0.45$, interference among $\lambda$ and $\tilde \lambda$ contributions
can render the model unobservable to current electroweak precision tests.

Being color triplets, leptoquarks can be pair-produced directly at hadron
colliders from $gg$ initial states, independent of their couplings to fermions.
The leptoquark that contributes  to $\bar B \rightarrow D^{(*)} \tau \nu$, $S_3$,
decays almost entirely to either $t\tau$ or $b\nu$ final states, with relative
branching fraction determined by the $\lambda$ and $\tilde{\lambda}$ couplings,
but with a branching fraction to $t\tau$ not less than $\approx 0.5$. Such
searches have been carried out in both channels by CMS~\cite{Khachatryan:2015bsa}.
Assuming equal branching fractions for a conservative bound, the mass
of the $S_3$ leptoquark is required to be $m_{S_3} \gtrsim 560\GeV$\@. Limits
from decay purely to the $b\nu$ final state can also be derived by reinterpreting
the bounds from sbottom searches~\cite{Khachatryan:2015wza,Aad:2013ija}, providing
comparable limits in areas of parameter space where they are applicable.

While the leading MFV interactions do not mediate a contribution to $b \to
s \nu\bar{\nu}$, the SM rate is sufficiently small that we also need to consider
effects subleading in the MFV expansion. The interactions in \eq{S113} refer
to the mass eigenstates only if the leptoquarks' couplings to different
down-type quarks do not mix. If such mixing occurs, it could induce large $b
\to s \nu\bar{\nu}$ transition. In a spurion analysis, we should consider any
number of spurion insertions. Since we need $y_b$ to be $\mathcal{O}(1)$, we
should include spurions of $Y_d$ at all orders~\cite{Kagan:2009bn}, and
the above conclusions should be checked after allowing such insertions. Thus,
the leptoquark mass matrix takes the form (again in the quark mass basis)
\begin{equation}
  \label{eq:Smfvmass}
  \begin{split}
    m_S^2 &= m_0^2\, (I + b Y_d^\dagger Y_d + \cdots) \\
      \Rightarrow
    m_{S_i}^2 &= m_0^2\, (1 + b y_{d_i}^2 + \cdots)\,.
  \end{split}
\end{equation}
This makes it clear that the leptoquarks do not mix to all orders in the MFV
expansion, while the third generation leptoquark can be split in
mass from the first two by a sizable amount.

Although the leptoquark mass matrix does not induce large $b \to s \nu\bar{\nu}$
transitions, strong constraints arise from interactions at next order in the MFV
expansion.\footnote{Due to the large value of $y_t$, terms to all orders
in $(Y_u Y_u^\dagger)^n$ should also be included in the expansion analogously
to the $(Y_d Y_d^\dagger)^n$ terms in \eq{Smfvmass}. Up to terms suppressed
by $m_c/m_t$, however, this only leads to a rescaling of the coefficient
in the first subleading term.} At leading order, \eq{S113} couples down-type
quarks to neutrinos, but as explained above, does not allow them to mix.
The next term, with three Yukawa spurion insertions,
\begin{equation}
  \delta\mathcal{L}' = \lambda' S Y_d^\dagger Y_u Y_u^\dagger\,
    \bar{q}^c_{L} i\tau_2 \ell^{\phantom c}_L \,,
\end{equation}
generates interactions of the form
\begin{equation}
  \label{eq:Smfvfcnc}
  \delta\mathcal{L}' = S_i\, \lambda'\, y^{\phantom 2}_{d_i} V^*_{ji}\, y_{u_j}^2 
  \left( \bar{u}^c_{Lj} \tau^{\phantom c}_L - V^{\phantom *}_{jk}\,\bar{d}^c_{Lk} \nu^{\phantom c}_L \right).
\end{equation}
This interaction generates $b \to s \nu\bar{\nu}$ decay,
to which the leading contribution is
\begin{equation}
  \mathcal{O}_{bs\nu\bar\nu} = -\frac{y_t^2 y_b^2\, \lambda\lambda'}{2m_{S_3}^2}\,
    V^*_{tb} V^{\phantom *}_{ts} \left(\bar{b}_L \gamma^\mu s_L\, \bar{\nu}_L \gamma_\mu \nu_L \right)\,,
\end{equation}
with no parametric suppression relative to $b \to c$.  Such transitions are
most strongly constrained at present by~\cite{Lees:2013kla}
\beq
  \mathcal{B}(B^+ \to K^+ \nu \bar{\nu}) < 1.6 \times 10^{-5}\,.
\eeq
Comparing the SM prediction with this bound~\cite{Altmannshofer:2009ma,
Ishiwata:2015cga} gives $\lambda'\lambda y_b^2/(2m_{S_3}^2) \lesssim
0.26/(1\TeV)^2$. If $C''_{S_R}$ in \eq{S113Bop} is to explain the values of
$R(D^{(*)})$, this can be rewritten as $\lambda'/\lambda \lesssim 0.1$.
Thus, a moderate suppression of the subleading coefficient of the MFV expansion
is necessary to avoid the $b \to s \nu\bar{\nu}$ constraint.

\subsubsection{Scalar leptoquarks --- lepton MFV}
\label{sec:sLQMFV}

We can extend the MFV framework to the lepton sector, and give the leptoquark
an appropriate charge under the $U(3)_L \times U(3)_e$ symmetry broken by the
lepton Yukawa couplings. Assigning the leptoquark to either of the lepton flavor
representations $(\arepthr,\repone)$ or $(\repone,\arepthr)$ ensures that either
the terms with $\lambda$ or $\tilde{\lambda}$, respectively, will couple to all
lepton generations uniformly. In the first case, the leptoquark contributes
as much to $B \to D^{(*)} l\bar{\nu}$ as to final states with a $\tau$,
up to effects of subleading spurion insertions, and $R(D^{(*)})$ cannot deviate
significantly from the SM prediction. In the second case, all couplings
generating corrections to $\bar B \to D^{(*)} l\bar{\nu}$ are suppressed
by Yukawas of either lighter quarks or leptons.

For this second case, the leading interaction terms take the form
\begin{align}
\label{eq:S11313}
  \delta\mathcal{L}
    &= S_{il}(\lambda {Y^*_d}_{ji} {Y_e}_{ml}\, \bar{q}^c_{Lj} i\tau_2 \ell^{\phantom c}_{Lm}
       + \tilde{\lambda} {Y^*_d}_{ji} {Y_u}_{jk}\,\bar{u}^c_{Rk} e^{\phantom c}_{Rl}) \nn \\
    &= S_{il} (\lambda y_{d_i} y_{e_l} V^*_{ji} \,\bar{u}^c_{Lj} \tau^{\phantom c}_{Ll}
     - \lambda y_{d_i} y_{e_l} \bar{d}^c_{Li} \nu^{\phantom c}_{Ll} \\
    &\qquad \quad + \tilde{\lambda} y_{d_i} y_{u_j} V^*_{ji}\, \bar{u}^c_{Rj} \tau^{\phantom c}_{Rl})\,. \nn
\end{align}
The leptoquarks coupling to lighter lepton generations introduce additional
constraints. Of those involving the $\lambda$ coupling, the leading constraint
comes from considering $D^0 \to \mu^+\mu^-$ decays, which receive a tree-level
contribution from $t$-channel leptoquark exchange. However, the contribution
to the branching ratio is suppressed by $\mathcal{O}(y_\mu^2)$ (in addition to
the already-present helicity suppression), making it two orders of magnitude
below current bounds~\cite{Aaij:2013cza} when consistent with the fit
to $R(D^{(*)})$. A contribution to $B_s \to \mu^+\mu^-$ is generated by a mixed
$W$--$S_{32}$ box diagram, but the resulting correction to the SM rate is
at the percent level.  A more severe constraint on $\tilde{\lambda}$ exists
from corrections to $R_\mu$, measured at LEP, due to the unsuppressed coupling
of the $S \bar{t}^c_R \ell_R$ vertex, providing a limit $\tilde{\lambda}/\lambda
\lesssim 0.8$.

The leptoquarks' coupling to light leptons mean that collider searches for first-
and second-generation leptoquarks become relevant. These bounds are less
constraining then the third-generation searches in \sec{sLQ}, however, since
they involve different leptoquark components than that contributing
to $R(D^{(*)})$. Here, the mass matrix takes the form
\begin{align}
  \label{eq:Slmfvmass}
    (m_{S}^2)_{ijlm} &= m_0^2\, (I + b {Y^*_d}_{ki} {Y_d}_{kj} + \cdots
                               + d {Y^*_e}_{kl} {Y_e}_{km} + \cdots) \nn \\
      \Rightarrow
    m_{S_{il}}^2 &= m_0^2\, (1 + b y_{d_i}^2 + d y_{e_l}^2 + \cdots)\,,
\end{align}
so the $S_{33}$ component can be lighter than the others by an $\mathcal{O}(1)$
amount without a sizable tuning. The fact that the bound on doubly-produced
first- and second-generation leptoquarks are approximately a factor of 2 more
stringent than those for the third-generation ones~\cite{CMS:2014qpa,CMS:zva}
does not then lead to any unavoidable additional constraints. Furthermore,
the additional leptoquarks do not lead to more stringent constraints from
$b \to s \nu\bar{\nu}$ decays, since all couplings to neutrinos other than
$\nu_\tau$ are Yukawa suppressed.

\subsubsection{Vector leptoquarks}
\label{sec:vLQ}

Consider a vector leptoquark, $U^\mu$, with $U(3)_Q \times U(3)_u \times U(3)_d$
charge $(\bm{1},\bm{1},\bm{3})$. If the leptoquark is an electroweak singlet,
the lowest order MFV contribution in the quark mass basis is
\begin{align}
\label{eq:Umfvint}
  \delta\mathcal{L}
    &= (\lambda\,\bar{q}_L Y_d \gamma_\mu \ell_L
       + \tilde{\lambda}\, \bar{d}_R \gamma_\mu e_R)\, U^\mu \\
    &= \big(\lambda y_{d_i} V_{ji} \,\bar{u}_{Lj} \gamma_\mu \nu_L
       + \lambda  y_{d_i} \bar{d}_{Li} \gamma_\mu \tau_L
    + \tilde{\lambda} \bar{d}_{Ri} \gamma_\mu \tau_R\big)\, U^\mu_i, \nn
\end{align}
while for the electroweak triplet, it is
\begin{align}
\label{eq:U3mfvint}
  \delta\mathcal{L}
    &= (\lambda\,\bar{q}_L Y_d \vec{\tau} \gamma_\mu \ell_L)\, \vec{U}^\mu \nn\\
    &= \big(\lambda y_{d_i} V_{ji} \,\bar{u}_{Lj} \gamma_\mu \nu_L
       + \lambda  y_{d_i} \bar{d}_{Li} \gamma_\mu \tau_L\big)\, U^{0\mu}_i\\
       &\quad + \frac{\lambda}{\sqrt{2}} \big(y_{d_i} \bar{d}_{Li} \gamma_\mu \nu_L\, U^{+\mu}_i
 	+ y_{d_i} V_{ji} \,\bar{u}_{Lj} \gamma_\mu \tau_L\, U^{-\mu}_i\big)\,. \nn
\end{align}
(Here the signs on $U$ do not indicate charges, but rather which component of
the triplet we refer to.)
The contribution to $b \to c$ decays comes from
\begin{equation}
\label{eq:Umfvop}
  \frac{C'_{V_L}}{\Lambda^2} =  \frac{V_{cb}}{m_{U_3}^2}\, \lambda^2 y_b^2\,, \qquad
  \frac{C'_{V_R}}{\Lambda^2} = \frac{V_{cb}}{m_{U_3}^2} \lambda\tilde{\lambda}\,y_b\,.
\end{equation}
As in the prior case, this leptoquark model requires $y_b = \mathcal{O}(1)$
to be able to explain the $R(D^{(*)})$ data with perturbative couplings.
As before, the electroweak triplet generates an operator with the wrong sign
for all values of $\lambda$. Using \tab{ops} and \fig{1Dfit}, we see that
a value of $\lambda y_b/m_{U_3} \approx (2.2 \pm 0.4)/\text{TeV}$ alone gives
a good fit.

If we consider giving leptonic flavor charges to the leptoquarks,
as in the scalar case, then the $(\arepthr,\repone)$ representation prevents
a deviation of $R(D^{(*)})$ from being generated as before.
The $(\repone,\arepthr)$ representation does not, but in this case, a tree-level
contribution to $B_s \to \mu^+\mu^-$~\cite{Aaij:2013aka,Chatrchyan:2013bka} is
generated at leading order in the MFV expansion. We estimate this to give
a constraint of $\lambda y_b/m_{U_3} \lesssim 1.0/\text{TeV}$, in significant
tension with the value required to fit the $B$ decay data. However, coupling
to purely $\tau$ can be enforced via a horizontal symmetry, as in \sec{sLQ},
and this is what we assume for the rest of the discussion.

The coupling to right-handed first-generation quarks is unsuppressed by any
Yukawa factors, and bounds on $\tau^+ \tau^-$ production from $Z'$ searches
again play a role, leading to a estimated limit on the right-handed
coupling of less than $\tilde{\lambda}/m_{U_1} \sim 0.12/\text{TeV}$, requiring
comparable suppression to that of the subleading coefficients of the MFV
expansion in the scalar case and for any contribution to $R(D^{*})$
from $\mathcal{O}'_{V_R}$ to be negligible. Integrating out the electroweak
singlet at tree level only generates four-fermion operators coupling neutrinos
to up-type quarks. The lack of data on $c \to u\nu\bar{\nu}$ transitions means
that subleading contributions in the MFV expansion to rare meson decays are
not a concern.

%As in the scalar case, we need to consider the possibility of generating sizable
%contributions to $b \to s \nu\bar{\nu}$ from additional spurion insertions. The
%leptoquark mass matrix remains diagonal. As in the scalar case,
%for the electroweak triplet vector leptoquark, the next  contribution to the MFV
%expansion of the interaction term, with three Yukawa spurion insertions,
%\begin{equation}
%  \delta\mathcal{L}' = \lambda'\,\bar{q}_L Y_u Y_u^\dagger Y_d\,
%  \vec{\tau} \gamma_\mu \ell_L \vec{U}^\mu ,
%\end{equation}
%leads to an operator of the form
%\begin{equation}
% \label{eq:Umfvfcnc}
% \mathcal{O}_{bs\nu\bar\nu} = \frac{V^*_{tb} V^{\phantom *}_{ts}}{m_{U_3}^2}\, y_t^2 y_b^2\,
%  \lambda'\lambda\, \left(\bar{b}_L \gamma^\mu s_L\, \bar{\nu}_L \gamma_\mu \nu_L \right).
%\end{equation}
%In this case, the bound is
%$\lambda'\lambda y_b^2/m_{U_3}^2 \lesssim 0.13/(1\TeV)^2$.
%If the operators in \eq{Umfvop} are to explain the
%$R(D^{(*)})$ data, this can be rewritten as $\lambda'/\lambda \lesssim 0.05$.
%The electroweak triplet case thus requires a suppression of
%the subleading coefficient of the MFV expansion to avoid violating
%the $b \to s \nu\bar{\nu}$ bound.  There is no such constraint for the
%electroweak singlet vector leptoquark, however, that model requires complex
%couplings to fit the $R(D^{(*)})$ data.

\begin{figure}[tb]
\includegraphics[width=\columnwidth]{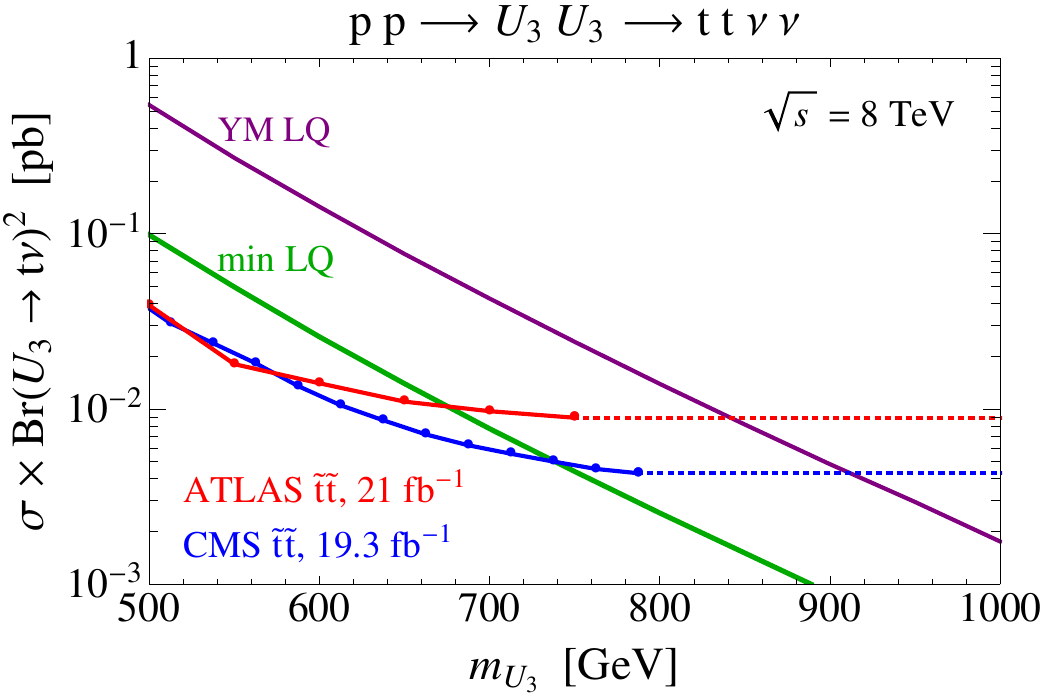}
\caption{Limits on $m_{U_3}$ from direct $pp \to U_3 U_3$ production.}
\label{fig:VLQstoplim}
\end{figure}

There are bounds on the leptoquarks produced directly through their gauge couplings via $gg
\to UU$ pair production. The leptoquarks can then decay to $b\tau$ or $t\nu$
pairs. Direct leptoquark searches exist for the first decay mode, while bounds
on the second mode can be derived by recasting stop searches. Vector leptoquark
production at hadron colliders is complicated by the fact that in the absence of
a UV completion, a marginal coupling not fixed by the minimal coupling
prescription is present. In addition to the interactions given by \eq{Umfvint},
the rest of the Lagrangian for the vector leptoquark is
\begin{equation}
  \label{eq:vLQdipole}
  \mathcal{L}_U = -\frac{1}{2}U^{\dagger}_{\mu\nu} U^{\mu\nu} + m_U^2 U^{\dagger}_\mu U^\mu 
   -i g_s \kappa\, (U^{\dagger}_\mu t^a U^{\phantom \dagger}_\nu)\, G^{\mu\nu}_a,
\end{equation}
up to higher-order operators suppressed by the cutoff. Here $U_{\mu\nu}
= D_\mu U_\nu - D_\nu U_\mu$, while an additional gauge-invariant ``dipole term''
beyond the minimal coupling prescription is also present. If the vector
leptoquark is a massive gauge boson from a spontaneously broken symmetry, then $\kappa = 1$, but in principle $\kappa$
can be arbitrary. Setting $\kappa = 0$ gives the minimal production cross section.
In all cases, a bound is produced directly on leptoquark masses, since the
leptoquarks are dominantly produced due to their gauge couplings, and not their
couplings to fermions. These are shown in \fig{VLQstoplim}\@. Since the
leptoquark direct production rate is higher than that of stops, data up to the leptoquark exclusion limit is not publicly
available (although stop searches could be recast for higher masses, which we leave for future work). In the figure we present a conservative extrapolation. 

Additional bounds come from meson mixing and precision electroweak measurements
of $Z \to \tau^+ \tau^-$. These were computed in Ref.~\cite{Eboli:1996hj}. The
interpretation of these is complicated by the fact that massive vector
leptoquarks are not in themselves UV-complete and yield logarithmic divergences
at loop level.  Similar to the scalar case, for the electroweak singlet vector
leptoquark, the next contribution to the MFV expansion of the interaction term,
with three Yukawa spurion insertions,
\begin{equation}
  \delta\mathcal{L}' = \lambda'\,\bar{q}_L Y_u Y_u^\dagger Y_d\, \gamma_\mu \ell_L U^\mu ,
\end{equation}
leads to an operator of the form
\begin{equation}
 \label{eq:Umfvfcnc}
 \mathcal{O}_{bs\tau\bar\tau} = \frac{V^*_{tb} V^{\phantom *}_{ts}}{m_{U_3}^2}\, y_t^2 y_b^2\,
  \lambda'\lambda\, \left(\bar{b}_L \gamma^\mu s_L\, \bar{\tau}_L \gamma_\mu \tau_L \right).
\end{equation}
The same interactions that lead to \eq{Umfvfcnc}, also give rise
to a quadratically divergent contribution to $B_{d,s}$ mixing
at one-loop level. As a conservative estimate, the contribution
of the finite part is then
\begin{equation}
  \label{eq:vLQMix}
  \mathcal{O}_{\Delta B = 2} = \frac{y_t^4 y_b^4 \lambda^{\prime2} \lambda^2}{4\pi^2 m_{U_3}^2}
    |V_{tb}^* V_{ts}|^2\, (\bar{b}_L \gamma^\mu s_L)^2.
\end{equation}
With current constraints~\cite{Charles:2013aka}, and again fixing
$\lambda^2/m_{U_3}^2$ to explain the observed excess, this implies
$\lambda^{\prime 2} \lesssim 0.1$, consistent with reasonable values of $\lambda$
without any greater tuning than already necessary. 

In the MFV framework, we thus find that two models are compatible with the data,
once mild tuning is allowed for subleading MFV couplings to evade
constraints from $b \to s\nu\bar{\nu}$: a scalar leptoquark in the
$(\bar{\bm{3}},\bm{1})_{1/3}$ and a vector leptoquark in the
$(\bm{3},\bm{1})_{2/3}$ representation of the SM gauge groups and in the
(anti)fundamental of the $U(3)_d$ flavor symmetry group. Contrary to assumptions
made elsewhere, although highly constrained, MFV physics can explain
the $R(D^{(*)})$ data.

\section{Summary and Future Signals}
\label{sec:conc}

We studied possible explanations of the observed enhancements of the $\bar B\to
D^{(*)}\tau\bar\nu$ rates compared to the SM predictions.  We identified which
higher dimension operators provide good fits to the data and determined their
required coefficients.  Since a sizable modification of a tree-level process in
the SM is required, the enhancements of $\bar B\to D^{(*)}\tau\bar\nu$ point to
a light, tree-level, mediator.  We found viable models, consistent with MFV,
where the mediator is a leptoquark.  While these models are consistent with
present flavor and LHC data, the low mass scale of the mediator, $m \lesssim 
1$~TeV, implies a variety of promising signals at future experiments. 

With more precise future data from Belle~II and LHCb, the necessary size of
flavor violation may decrease while maintaining a  high significance for NP\@. 
In this paper, we focused on NP that can accommodate the central values of the
current measurements, but in reality NP may have a heavier mass scale,
alleviating possible tensions with present constraints.  If the magnitude of the
deviation from the SM decreases while its experimental significance does not, it
will become important to control theoretical uncertainties as well as possible.
It would likely be advantageous to consider ratios~\cite{Tanaka:1994ay} in which
the range of $q^2$ integration is the same in the numerator and the denominator,
\beq
\widetilde R(D^{(*)}) = \frac{\ds \int_{m_\tau^2}^{(m_B-m_{D^{(*)}})^2} 
\frac{d\Gamma(B \to D^{(*)} \tau\bar\nu)}{d q^2}\, dq^2 }
{\ds \int_{m_\tau^2}^{(m_B-m_{D^{(*)}})^2} 
\frac{d\Gamma(B \to D^{(*)} l\bar\nu)}{d q^2}\, dq^2 }\,,
\eeq
which have smaller theoretical uncertainties than $R(D^{(*)})$.  The reason is
that including the $0 < q^2 < m_\tau^2$ region in the denominator simply dilutes
this ratio and the sensitivity to new physics effects, and also because the
uncertainties of the semileptonic form factors increase at smaller $q^2$ (larger
recoil).

We conclude by enumerating future experimental measurements and new physics
search channels that can test for models motivated by the measured $\bar B\to
D^{(*)}\tau\bar\nu$ rates. \begin{enumerate}\itemsep 0pt

\item More precise data on $\bar B\to D^{(*)}\tau\nu$, including measurements of
differential distributions and their correlations.  These would also help
discriminate NP from the SM, and between different NP models.

\item $D\to \pi\nu\bar\nu$ could be near the $10^{-5}$ level, which may be
observable in BES~III~\cite{BESprivate}.

\item In some scenarios a nonresonant contribution to $t \to c \tau^+ \tau^-$ is
possible. While current $t \to cZ$ searches~\cite{Aad:2012ij,Chatrchyan:2013nwa} focus on $m_{\ell^+\ell^-}$
consistent with $m_Z$, there could also be a contribution that is inconsistent with the $Z$ lineshape.

\item A NP contribution to $t\to b\tau\bar\nu$ of a few times
$10^{-3}\,\Gamma_t$ is possible.  Noting that the $\tau\bar\nu$ do not come from a $W$ decay in the NP contribution, a high transverse mass tail, $m_T(\tau, \slashed{E}_T) > m_W$, can be searched for in semileptonic $t \to \tau$ decays.

\item In this paper we neglected $CP$ violation.  A simple experimental test is
to constrain the difference between $B^+$ and $B^-$ rates to
$D^{(*)}\tau\bar\nu$; a measurable effect would be a clear sign of new physics.

\item Models that contribute to $\bar B\to D^{(*)}\tau\nu$ can modify $Z \rightarrow \tau^+ \tau^-$ at the order of present constraints.  Therefore, future precision $Z$-pole measurements may detect deviations from the SM.

\end{enumerate}

Finally, we note that there may be viable models in which the $B\to
D^{(*)}\tau\bar\nu$ rates are given by the SM, but $B\to D^{(*)} l \bar\nu$
($l=e,\mu$) are suppressed by interference between NP and the SM~\cite{future}.

\begin{acknowledgments}

We are grateful to Florian Bernlochner, Manuel Franco Sevilla, and Vera Luth
for useful discussions about the BaBar analysis~\cite{Lees:2012xj}, and we
thank Cliff Cheung, Lawrence Hall, Yossi Nir, Yasunori Nomura, Michele Papucci,
Gilad Perez, and Marcello Rotondo for helpful conversations. We thank the authors
of Ref.~\cite{Aloni:2017ixa} for bringing a sign mistake to our attention
in the vector leptoquark discussion in a prior version of this paper, which
alters which vector leptoquark is viable and some of the resulting constraints.
MF and ZL thank the hospitality of the Aspen Center for Physics, supported 
by the NSF Grant No.~PHY-1066293. MF thanks Lawrence Berkeley National Laboratory
and ZL thanks the CERN theory group for hospitality, while parts of this work
were completed. This work was supported in part by the Director, Office of Science,
Office of High Energy Physics of the U.S.\ Department of Energy under contract
DE-AC02-05CH11231, the U.S.\ Department of Energy under grant DE-SC003916 and
the National Science Foundation under grant No. PHY-125872.

\end{acknowledgments}

\appendix
\section{\boldmath Comment on $U(2)^3$ scenarios}
\label{sec:U23}

We presented several MFV models capable of describing the $R(D^{(*)})$ data
while also coupling to other quark generations consistently with all
constraints.  As detailed earlier, extending the MFV framework to the lepton
sector in these scenarios is not always possible, and so we have assumed 
horizontal symmetries governing the structure of lepton flavor where necessary.
However, it is possible to treat quark and lepton flavor uniformly in all
scenarios, at the cost of a slightly less predictive flavor framework.

A minimal self-consistent extension of the MFV framework exists in the form
of the $U(2)^3$ flavor symmetry approach~\cite{Barbieri:2011ci, Barbieri:2012bh}.
Here, the UV physics is only assumed to preserve a $U(2)_Q \times U(2)_u
\times U(2)_d$ symmetry for two generations in the quark sector, which is
closely aligned with the first two generations of the SM\@. Three spurions are
now necessary to minimally characterize the quark sector flavor symmetry
breaking, here denoted as
\begin{equation}
  \Delta_u \sim (\bm{2}, \bar{\bm{2}}, \bm{1})\,, \quad
  \Delta_d \sim (\bm{2}, \bm{1}, \bar{\bm{2}})\,, \quad 
  \bm{V} \sim (\bm{2}, \bm{1}, \bm{1})\,,
\end{equation}
while the quark fields decompose as
\begin{equation}
\bm{q}_L \sim (2,1,1)\,, \quad
\bm{u}_R \sim (1,2,1)\,, \quad 
\bm{d}_R \sim (1,1,2)\,, 
\end{equation}
and the singlet third generation fields are $q_{3L}$, $t_R$, and $b_R$.
The extension to the lepton sector, with an additional $U(2)_L^2 \times U(2)_e$
symmetry, is straightforward, and only a single spurion, $\Delta_e \sim
(\bm{2}, \bar{\bm{2}})$ is needed (ignoring $m_\nu$, which requires additional but much smaller spurions). Unlike the case
of MFV, none of these spurions have $\mathcal{O}(1)$ components, so higher-order
spurion insertions are highly suppressed.

Coupling predominantly to third generation quarks and leptons can now be
imposed by demanding that the mediator is a singlet under all flavor symmetries,
as long as currents of light quarks and leptons cannot be written as flavor
singlets.

As an illustrative example, we consider a scalar leptoquark. Due to the $V$
spurion, this leptoquark can couple to all quark generations. The interaction
terms are
\begin{multline}
\delta\mathcal{L}
  = S\, \big(\lambda_1\, \bar{q}^c_{3L} i\tau_2 \ell_{3L}
    + \lambda_2 \bm{V}\,\bar{\bm{q}}^c i\tau_2 \ell_{3L} \\
    + \tilde{\lambda}_1\,\bar{t}^c_{R} \tau_R
    + \tilde{\lambda}_2 \bm{V}^\dagger \Delta_u \bar{\bm{u}}^c \tau_R\big)\,.
\end{multline}
Coupling to the other lepton generations is impossible, since no combination
of spurions can absorb their flavor charge. The contribution
due to terms proportional to $\tilde{\lambda}_2$ can be neglected, as it
always comes with multiple spurion suppression factors. Then the leading
contribution by spurion counting to $b \to c$ transitions is given by
\begin{equation}
\label{eq:S111Bop}
  \frac{C''_{S_R}}{\Lambda^2} = \frac{V^*_{cb}}{m_S^2}\,
  \lambda_\alpha \lambda_1\,, \qquad
  \frac{C''_{S_L}}{\Lambda^2} = \frac{V^*_{cb}}{m_S^2}\,
  \lambda_\alpha \tilde{\lambda}_1\, y_c\,.
\end{equation}
Here $\lambda_\alpha$ is in the range $(\lambda_1,\, \lambda_2)$ with the
precise value set by the coefficients of $\bm{V}$ in up- and down-type Yukawa
couplings. In the limit of $\lambda_1 = \lambda_2$ (and $\tilde{\lambda}_1 =
\tilde{\lambda}_2$ if we include higher order terms in our Yukawa
diagonalization), we recover the contribution of the third generation leptoquark
of an MFV scenario. This is as we would expect. If we were to add a leptoquark
in, say, the $(\repone, \repone, \bm{2})$ representation to the model above,
setting all the suprion coefficients of terms with $\ell_{3L}/\bm{\ell}$
to $\lambda$ and those with $\tau_{R}/\bm{e}_R$ to $\tilde{\lambda}$,
we precisely recover the particle content and spurion structure of the
$(\repone,\repone,\arepthr)$ MFV model.

The analysis of the constraints follows the MFV case. The treatment of
contributions to $b \to s \nu\bar{\nu}$ is different, however. The term
$\lambda_\alpha$ can be written, ignoring phases and using
$|V_{cb}| \ll 1$,
\begin{equation}
\label{eq:bcexact}
  \lambda_\alpha = \frac{\lambda_2-x_t \lambda_1}{x_b - x_t}\,,
\end{equation}
where $x_{b,t}$ are free parameters defined in Ref.~\cite{Barbieri:2011ci}.
However, in this case contributions to $b \to s \nu\bar{\nu}$ are generated
by the same interactions. Similar to \eq{bcexact} above,
the coefficient of this interaction can be written as
\begin{equation}
\label{eq:bsexact}
  \frac{V_{cb}}{2}\, \lambda_1 \frac{\lambda_2-x_b \lambda_1}{x_b - x_t},
\end{equation}
so one can reduce the rate of the interaction by tuning $x_b \approx
\lambda_2/\lambda_1$ without affecting the $b \to c \tau \bar{\nu}$ rate.
A slight tuning is then again necessary to avoid this constraint, although
arising in a different set of couplings.

\end{document}